\documentclass{IEE-CTA}

\usepackage[utf8]{inputenc}
\usepackage{graphicx}
\usepackage{todonotes}
\usepackage{mathtools}
\usepackage{pgfplots}
\usepackage[utf8]{inputenc}
\usepackage{amsthm}
\usepackage{amsmath}
\usepackage{indentfirst}
\usepackage{upgreek}
\usepackage{url}
\usepackage{algorithm}
\usepackage{algpseudocode}
\usepackage{algorithmicx}

{}
{}
{}

\begin{document}

\supertitle{This paper is a preprint of a paper submitted to IET Communications journal. If accepted, the copy of record will be available at the IET Digital Library}

\title{Transmit power optimization in cellular networks with nomadic base stations}

\author{\au{Krzysztof Grochla$^{1\corr}$}, \au{Mariusz Slabicki$^{2}$}}

\address{\add{1}{Institute of Theoretical and Applied Informatics, Polish Academy of Sciences, Baltycka 5, Gliwice, Poland}
\add{2}{Nokia Solutions and Networks, Szybowcowa 2, Wroclaw, Poland}
\email{kgrochla@iitis.pl}}


\begin{abstract}
The increasing demand for cellular network capacity can be mitigated through the installation of nomadic eNodeB, which serve a temporal increase of traffic volume in specific area. When nomadic cells are deployed, the transmission power of neighbor base stations needs to be optimized to limit the inter-cell interferences. We analyze the problem of neighborhood selection for the optimization, to define what part of the networks needs to be reconfigured when new base station is added. We evaluate the iterative approach, with increasing range of neighboring cells being reconfigured and propose a novel, sampling based local TX power reconfiguration method, which is evaluated by a numerical model in both regular (honeycomb) topology and in realistic topology reflecting locations of cells in a city. The analysis confirms that the proposed algorithm allows to select very few neighboring cells which need to be reconfigured (in majority of the cases less than 10 cells) and achieve similar efficacy as global optimization, with total network throughput different by less than 1\% comparing to the global optimization.  

\end{abstract}

\maketitle

\section{Introduction}\label{sec1}

The cellular network load depends on the amount of traffic transmitted to each client and is determined by spatial distribution of the users. The density of users and user terminals (UEs) changes in time and depends e.g. on the popularity of different locations or urban traffic patterns. To adjust the possible network throughput to users' demand, cellular operators build heterogeneous network (Het-Net) with variable cell sizes. The introduction of small cells in 4G and 5G systems allows to adapt the network topology to the demand for bandwidth in a given location. Installation of small cells is an enabler in reaching the IMT 2020 goals to increase the throughput of the network by 1000 times. The cost of a single cell decreases thanks to the development of low-cost small base stations, what allows to install them in multiple locations showing high traffic demand, leading to higher spatial reuse of the available frequencies. As a result, the throughput available to clients on a cell area, represented in the Area Traffic Capacity metric \cite{itu20202083} is increased in areas with multiple small cells. The cellular network is subject of constant changes, when new cells are added in the locations where the demand bandwidth grows.

The traditional model of radio planning relies on RF engineers to carry out measurement, RF surveys, and drive tests to determine the transmit power (TX power), antenna azimuth, and tilt. In particular, the TX power has very high influence on the inter-cell interferences and on the network coverage\cite{liu2019time}, because the LTE and 5G use the same frequencies in all cells. During the lifetime of the network these parameters need to be adapted e.g. when a new cell is added in the near-by location. For a network consisting of large number of small cells this is not feasible, as the cost of the RF planning and reconfigurations may be a significant part of operational costs. Moreover, the modern cellular networks become more dynamic, with new base stations added to the network in response to increased load to maximize the resource reuse.
The deployment of small cells is partially unplanned and makes the network topology irregular and hard to manage. The small cells may be flexibly turned on or off according to traffic requirements or new nomadic eNodeB may be installed, which makes the topology of the network more dynamic. As the complexity of the networks grows and the changes are frequent, it becomes very difficult for the operator to manually adjust the configuration of the base stations to minimize the inter-cell interferences, and maximize the throughput offered to the clients. 

The RF planning is hard to implement in the nomadic base station use case, when a new cell, or a small set of cells is added to the network to serve temporary traffic increase e.g. in response to natural disaster or during sport events. The small cells can be deployed within hours, but the network configuration must be adapted to limit the transmission power of the adjacent eNodeBs. The automation of the configuration of the eNodeB has been standardized by the 3GPP as an element of the Coverage and Capacity Optimization (CCO) function of the self-optimized network (SON). The goal of CCO functions is to provide optimal coverage and maximize the network capacity  \cite{3gpp.36.902}. The CCO should configure the network so the users can establish and maintain connections with acceptable or default service quality and must take the impact on network capacity into account. The CCO is realized through the control of transmit power, antenna tilt and antenna azimuth. In this work we concentrate on the control of transmit power. 

The contribution of our work is the analysis of the problem of the selection of optimal set of cells which needs to be reconfigured in response to change in the network topology (e.g. the addition of a base station) or in response to the change in the traffic demand density. In our previous work \cite{slabicki_grochla} we have shown that the local approach to the TX power selection provides comparable results to the problem of TX power management cellular networks, especially including small cells, providing much lower cost of the reconfiguration and while being less computation intensive. However, it is an open question what part of the network needs to be reconfigured when a new base station is added. In this work we extend the proposed method to automatically select the neighborhood that should be considered during optimization. 

The selection of the part of the network which needs to be reconfigured in response to the change of the network load or to addition of a base station in a specific location has no obvious solution. Using a map an experienced operator may manually indicate the list of cells which are subject to the changes in the level of the interferences and are potential candidates for the reconfiguration. Nevertheless, the change of the parameters of one base station may propagate through the network and require some adaptations also in parts of the network which are not within direct neighborhood. To the best of our knowledge, this problem has not been solved yet. We analyze how effective is the selection of the part of the network for the reconfiguration by iterative increase of reconfiguration range (described in section \ref{sec:iterative}) and prpose a sampling based reconfiguration algorithm (described in the section \ref{sec:sampling}).

The rest of the paper is organized as follows: in section \ref{literature_review} we present the literature review, in section 3 we describe the network model used for the analysis of power management algorithms, in section \ref{sec:TX_power_selection} we show the problem of transmission power optimization. Next we describe  iterative increase of reconfiguration range in section \ref{sec:iterative} and we propose a method for the selection of the range of neighbors which are subject to reconfiguration when new nomadic base station is added: sampling based reconfiguration in section \ref{sec:sampling}. We finish the article with a short conclusions in section \ref{sec:conclusions}.

\section{Literature review}
\label{literature_review}

The cell coverage and cell capacity is significantly influenced by the appropriate selection of transmission power, what has been shown e.g. in \cite{yilmaz2009analysis} where different network scenarios and various combinations of antenna parameter configurations have been studied in terms of signal to interference plus noise ratio (SINR) performance and cell throughput. The selection of the eNodeB configuration to maximize the traffic area capacity is a classical optimization problem. Most of the solutions proposed in the literature are solutions based on optimization methods applied to the whole network \cite{slabicki:aliu2013survey}. Methods such as simulated annealing \cite{balasubramanya2016simulated} \cite{slabicki:temesvary2009wireless}, Integer Linear Programming (ILP)  \cite{slabicki:lopez2011optimization}, fuzzy neural networks \cite{fan2014self} or game theory \cite{sen2012coverage} have been used to search for the optimal solution. The amount of inter-cell interferences can be also minimized by the use of scheduling of physical resource blocks and by fractional frequency reuse schemes \cite{aghababaiyan2017qos}, however this doesn't eliminate the need of optimization of transmission power selection. In the work \cite{buenestado2017self} the power planning is formulated as a large-scale nonseparable multiobjective optimization problem and an algorithm is proposed to improve the overall network spectral efficiency in the downlink by reducing the transmit power of specific cells to eliminate interference. However the global approach to the problem of coverage and capacity optimization neglects the fact that changing the base station configuration (including the transmit power) is costly because it may influence the ongoing transmissions in all the cells which are subject to reconfiguration. 

Currently, due to the decrease of base station costs, we are observing a change in network control paradigm. Traditionally, operator decided about where and when a base station has to be located during network planning and performed a large scale optimization \cite{chang2017energy}. The ability to decide about base stations placement is either shifted towards network users, which may install their own base station (femtocell), to cover the network traffic in small area (e.g. a home)  \cite{slabicki:raheem2013fixed}, or it is executed by the operator in response to change in network load through the addition of a nomadic cell \cite{bulakci2015towards}. A special case of dynamic cell addition is to use nomadic base stations, which may be added to network only when operator forecasts significant increase of the expected demand for network services in some area (i.e. during event which is popular among people). The nomadic eNodeB has been standardized by the 3GPP \cite{3gpp.33.897} as a deployable system which has the capability to provide radio access, local IP connectivity and public safety services to UEs. A more recent work by Bulakci et al. \cite{bulakci2017ran} proposes to further extend this concept into vehicular nomadic nodes (VNNs) which are low-power access nodes with wireless self-backhaul that can be activated temporarily to provide additional system capacity and/or coverage on demand.

Base stations added in an uncontrolled manner may cause decrease in Quality of Service due to the inter-cell interferences \cite{liu2019time}. Newly added nomadic base stations or femtocells may require reconfiguration of neighbor base stations, which should be done quickly and efficiently. Thus it is important to develop a method to automate the adaptation of configuration within the area of the current network which is influenced by the interferences of the newly added femtocell or nomadic eNodeB. The idea of reconfiguration of part of the network in the vicinity of the newly added nomadic cell or femtocell has been discussed in a few papers. 
Authors in \cite{slabicki:eisenblatter2011self} propose Soft Integration concept that reconfigures the added base station and base stations which can be seen as direct neighbors. However the paper does not provide analysis of the performance of the proposed concept and there is little said on the problem of neighborhood selection. A similar idea is described in \cite{slabicki:sanneck2010context}, where a Dynamic Radio Configuration framework is proposed. The authors of this paper prepared classification of configuration parameters and propose to differentiate the reconfiguration parameters depending on class. As in the previous article the paper fails to provide the validation of proposed methods. Ge, Jin and Leung propose to merge the power allocation problem with opportunistic user scheduling \cite{ge2017joint} and develop an algorithm which solves the resource allocation problem via two stages, user scheduling by the BS and power allocation by each user. The proposed scheme enables near-optimal selection for instantaneous transmit power of each user, but doesn't take into account the influence of the transmission power in neighboring cells. 

Two articles by Wang and Shen: \cite{wang2017small} and \cite{8403499} describe how to use the stochastic bandit theory to address the small base station (SBS) transmit power assignment problem. To guarantee good performance when the prior knowledge is insufficient the authors establish an algorithm that exploits the correlation structure to incorporate the performance correlation among similar power values. The simulations analysis given shows that the algorithm allows to decrease the packet loss ratio in sample scenarios, but the analysis of the range of configuration changes is oriented on reduction of the complexity of the problem, and there is no analysis of what part of the existing network should be reconfigured.

\section{System model}
\label{sec:system_model}
\begin{figure}
	\centering       
    \includegraphics[width=\columnwidth]{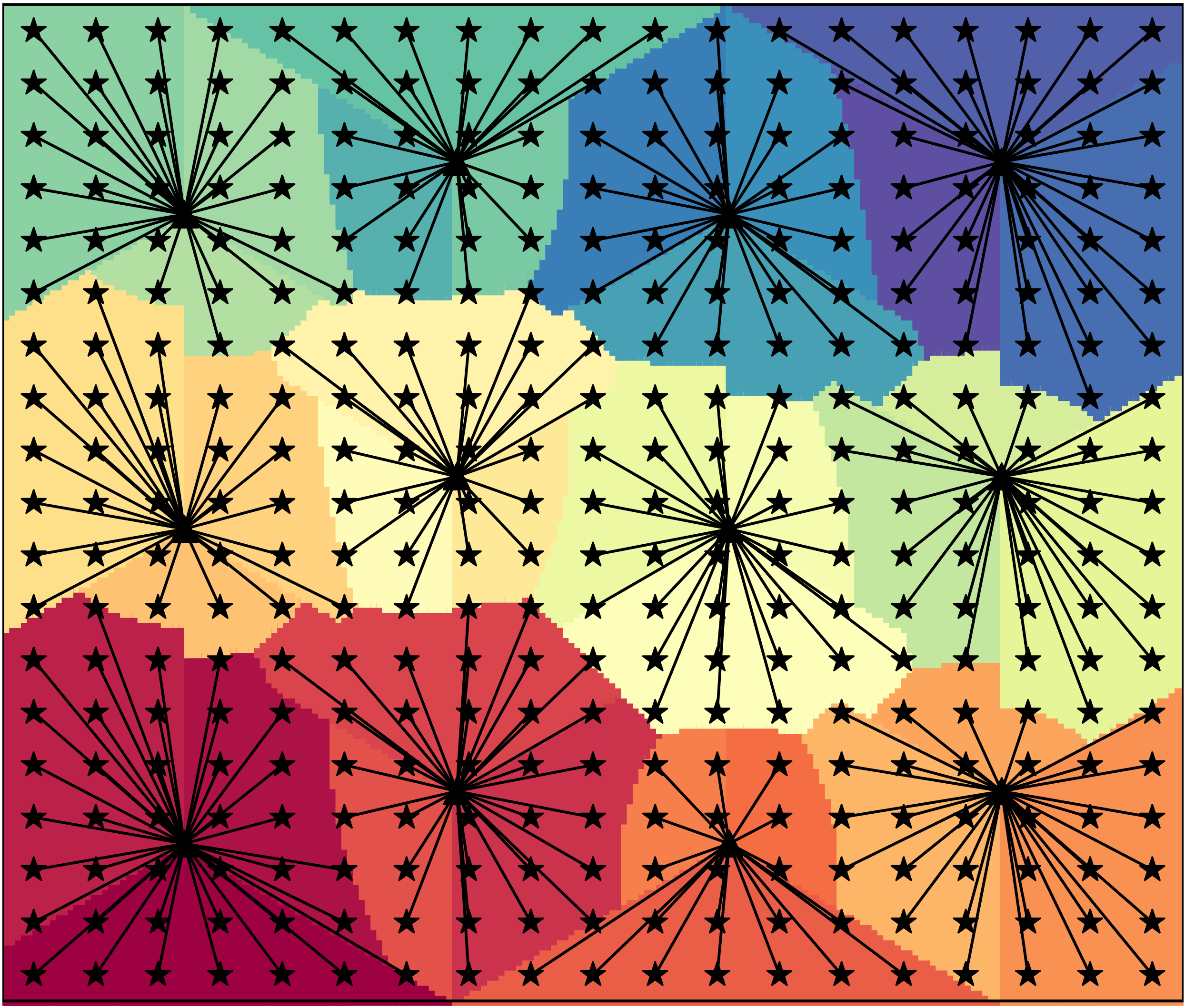}
    \caption{Reference network}
    \label{fig:local_referential_network}
\end{figure}

We consider a network consists of a set of $User Entity$ devices ($UE$) and a set of $Base Stations$ ($BS$). Each of those devices is located in a two-dimensional space, and have defined ($x, y$) position. All UEs and BSes are located in the rectangular area, which size is described as ($X, Y$). An example network is shown on the Figure \ref{fig:local_referential_network}. 
$BS$ devices have defined $Transmission Power$ parameter, which defines what is the energy radiated by the antenna. Every $BS$ also has a parameter $angle$, which is related with the direction of the centre of the radiated beam. The width of the beam is equal 120 deg. To assign $UE$ to $BS$ we use $best-SNR$ heuristic, which is widely used in wireless systems \cite{ali2011understanding}. The $best-SNR$ heuristic assigns a $UE$ to the $BS$ with highest $SINR$. The $SINR$ also defines what is the maximal possible throughput achievable by a $UE$. 

To determine SINR level for a UE connected to a particular eNodeB and taking into account the interferences from other cells, we use the following formula: 
\begin{equation}
\label{slabicki:sinr_eq}
SINR_{i} = \frac{P_{tx} - PL}{\sum_{BS_{vis}}(P_{tx}-PL) + N}
\end{equation}

where $P_{tx}$ is a transmission power level of the BS -- in numerator it is $P_{tx}$ of the BS to which an UE is connected, in denumerator we consider $P_{tx}$ of all eNodeBs that are visible from the UE. The $N$ value is a noise level and the $PL$ is the path loss calculated according to the SUI radio signal propagation model \cite{slabicki:erceg1999empirically}.




The maximum throughput for a given client is determined by the modulation and coding scheme (MCS) in a given location. The MCS is selected as a function of the signal quality indicator, which is calculated based on an average SINR at a given location. The SINR thresholds for the Additive white Gaussian noise (AWGN) case have been determined for the different MCSs with the corresponding efficiency ${\eta}_{i}$ according to the following model, presented earlier in \cite{giambene2014resource}:
\begin{equation} \label{GrindEQ__1_}
\begin{split}
\eta _{i} =\log _{2} \left(1+\frac{SINR_{i} }{\Gamma } \right)  
\end{split}
\end{equation}

where $\Gamma =-\frac{2}{3} \ln \left(5\times BER\right)$  and  BER = 0.00005 and where the efficiency of the \textit{i}-th MCS \textit{${\eta}_{i}$} can be determined on the basis of the data in Table \ref{table:mcs} according to the following formula:
\begin{equation} \label{GrindEQ__2_}
\eta _{i} =r_{i} \log _{2} \left(M_{i} \right)\quad \Rightarrow \quad 2^{\eta _{i} } =M_{i}^{r_{i} }
\end{equation}

\begin{table}
  \centering
  \caption{CQIs and SINR thresholds for AWGN channel conditions.}
\begin{tabular}{|p{0.3in}|p{0.5in}|p{0.6in}|p{0.5in}|p{0.6in}|} \hline
\textbf{\newline \textit{CQI${}_{i}$}} & \textbf{\newline Modula- tions} & \textbf{\newline Code rate \textit{r${}_{i}$}} & \textbf{\newline Modula- tion size, \textit{M${}_{i}$}} & \textbf{SINR thresholds, \textit{SINR${}_{i}$} (AWGN)} \\ \hline
1 & QPSK & 78/1024 & 4 & $-$2.1054 \\ \hline
2 & QPSK & 120/1024 & 4 & $-$0.1083 \\ \hline
3 & QPSK & 193/1024 & 4 & 2.1776 \\ \hline
4 & QPSK & 308/1024 & 4 & 4.5647 \\ \hline
5 & QPSK & 449/1024 & 4 & 6.6514 \\ \hline
6 & QPSK & 602/1024 & 4 & 8.4275 \\ \hline
7 & 16QAM & 378/1024 & 16 & 9.9379 \\ \hline
8 & 16QAM & 490/1024 & 16 & 11.8495 \\ \hline
9 & 16QAM & 616/1024 & 16 & 13.7624 \\ \hline
10 & 64QAM & 466/1024 & 64 & 14.9370 \\ \hline
11 & 64QAM & 567/1024 & 64 & 16.9703 \\ \hline
12 & 64QAM & 666/1024 & 64 & 18.8734 \\ \hline
13 & 64QAM & 772/1024 & 64 & 20.8506 \\ \hline
14 & 64QAM & 873/1024 & 64 & 22.6980 \\ \hline
15 & 64QAM & 948/1024 & 64 & 24.0546 \\ \hline
\end{tabular}
\label{table:mcs}
\end{table}

The throughput calculated according to the above described model can be used by a single client if it is solely using the whole available transmission time. When multiple {\it UEs} are sharing the radio resources of a given cell, the scheduler algorithm in the BS allocates the transmission time for each of the {\it UEs}. Assuming the Round Robin scheduler is implemented, the $\frac{1}{n}$ of the throughput is allocated for each of the UEs connected to the given BS, where $n$ is the number of {\it UEs} in a cell. As a consequence, the throughput available per UE may be calculated as the maximum throughput for the UEs divided by the number of UEs within the cell.

\begin{figure}
  \centering
  \includegraphics[width=0.95\linewidth]{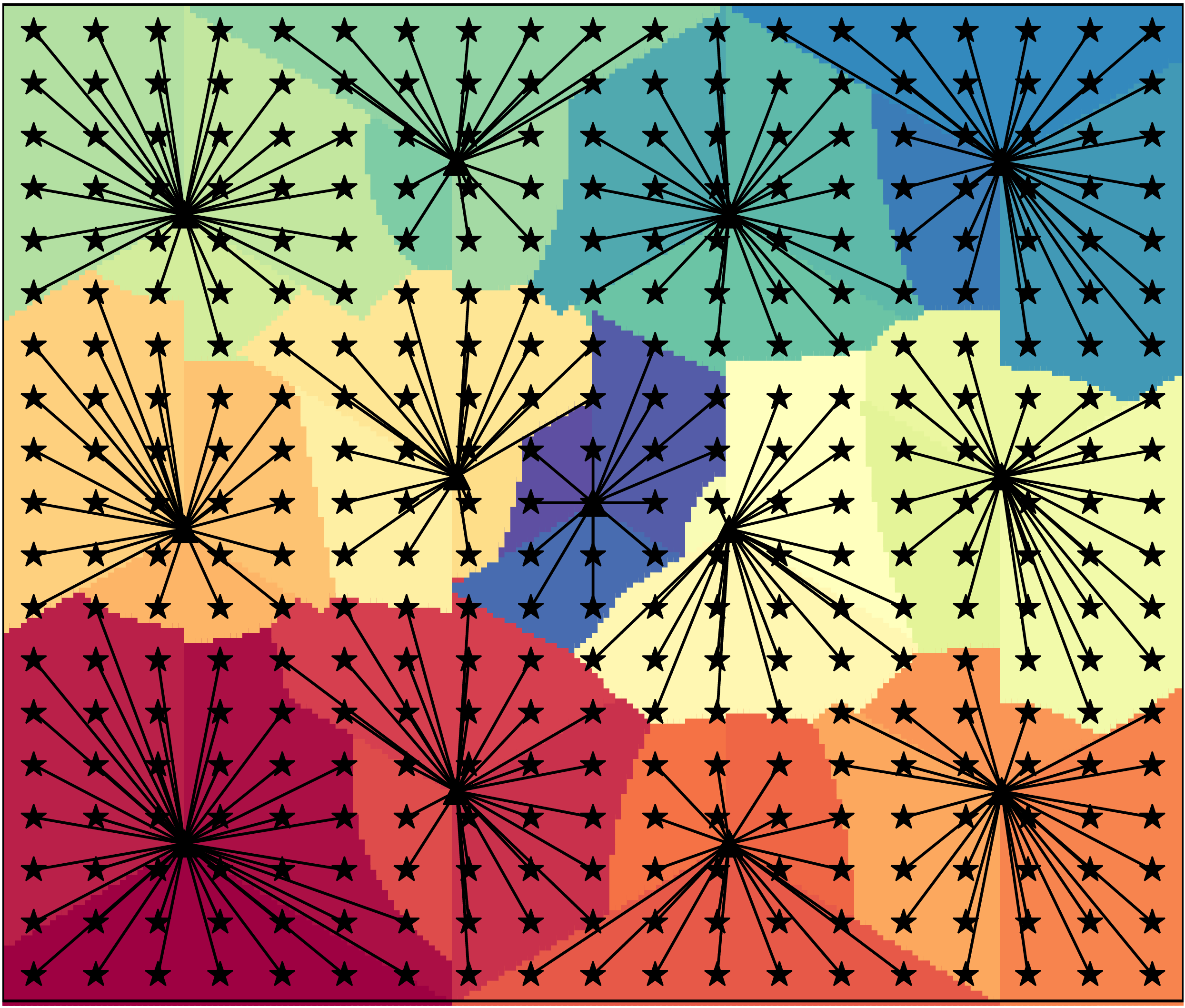}
  \caption{Network with added BS with TX Power selected by global optimization. (source: \cite{slabicki2015local})}
  \label{fig:add_global}
\end{figure}



We use the total network throughput as the efficiency metric of the network and the quality measure for the given base stations' configuration. We define the total network throughput as the sum of the offered throughput in all the cells in the network. In other words, the total throughput can be seen as an integral of Area Traffic Capacity \cite{itu20202083} over all the cells in the network. To evaluate particular configuration of the transmit power, the total network throughput is calculated as the sum of throughput of all UEs. 

\section{Transmit power selection}
\label{sec:TX_power_selection}



In our previous works: \cite{slabicki:slabicki2014automatic} and \cite{slabicki2015local} we have shown that the maximization of total throughput offered to the clients allows to correctly solve the problem of optimal selection of the transmit power in LTE networks. We use it as a reference for the analysis of the problem of the selection of reconfiguration range in response to the addition of a new nomadic cell. The sample configuration of TX power and coverage calculated as an output of a global optimization is presented on Figure \ref{fig:local_referential_network}. It uses the model of LTE network presented in section \ref{sec:system_model}.  The end nodes are placed in a regular grid within a rectangle 7498x6327 meters. The following parameters have been used in the evaluation number of cells: 36, number of UEs: 361. The colors denote boundaries of the cells, namely the area for which a SINR of a given BS is highest. The TX Power of each of the BS is selected by using Simple Genetic Algorithm, with the total throughput (sum of throughput offered to all clients) used as the goal function of the optimization.

Next we present the scenario of extension of the network by adding a nomadic eNodeB: a new base station is added in the central point of the network. When the global optimization of TX power was rerun after the adding of a new cell, the network is able to adapt to the new topology, but practically all the cells need to be reconfigured \ref{fig:add_global}.




\subsection{Local reconfiguration method}

The global optimization of the TX power in the whole network is costly, as there is high probability that even a small change in initial optimization conditions may lead of reconfigure of large part of the network. Thus it is important for a base station addition use case to limit the range of reconfiguration to some area in the vicinity of the location of nomadic eNodeB. In \cite{slabicki2015local} we have proposed a local reconfiguration method, which allows to define the range of cells which are subject of the TX Power optimization, without the changes in the remaining part of the network.

The following scenario was used to evaluate the reconfiguration method:
\begin{itemize}
\item find a power configuration for reference network,
\item add new cell with three sector antennas in central point of the network and run (independently) two algorithms:
\begin{itemize}
\item global reconfiguration algorithm (presented in Algorithm \ref{alg:global})
\item local reconfiguration algorithm
\end{itemize}
\item calculate metrics for each reconfiguration method.
\end{itemize}

Above method might be easily applied to cellular network of different topologies and geographic distribution of base stations. The optimization procedure has been executed 50 times to achieve statistically reliable results. We have used PyGMO library \cite{slabicki:izzo2012pygmo} with Simple Genetic Algorithm (SGA), configured to use 100 generations (this size was selected experimentally, to the level after which the increase of number of generations did not improve the results). The SGA method was selected as an optimization method applicable to the non-linear problems. 

\begin{algorithm}
  \caption{Global reconfiguration algorithm \cite{slabicki2015local}}
  \label{alg:global}
  \begin{algorithmic}
  \ForAll{user equipment}
    \State \Call{connect to the nearest BS}{}
  \EndFor
  \State \Call{run the metaheuristic simple genetic optimization}{}
  \ForAll{user equipment}
  \State \Call{connect to the BS with the highest SINR}{}
\EndFor
  \end{algorithmic}
  \end{algorithm}

Considering that the global method might results in complex reconfiguration behaviour of a network, in \cite{slabicki_grochla} the local approach to reconfiguration has been considered. The local method was based on similar procedure as in the global optimization, with a optimization limits set to a predefined list of BSes than can be reconfigured by the algorithm. A manually selected subset of base stations set e.g. to the 4 nearest BS locations provides proper results, the manual definition of the size and range of the neighborhood is not effective for large scale networks.

\section{Iterative increase of reconfiguration range}
\label{sec:iterative}

\begin{figure}
	\centering    
    \includegraphics[width=0.9\linewidth]{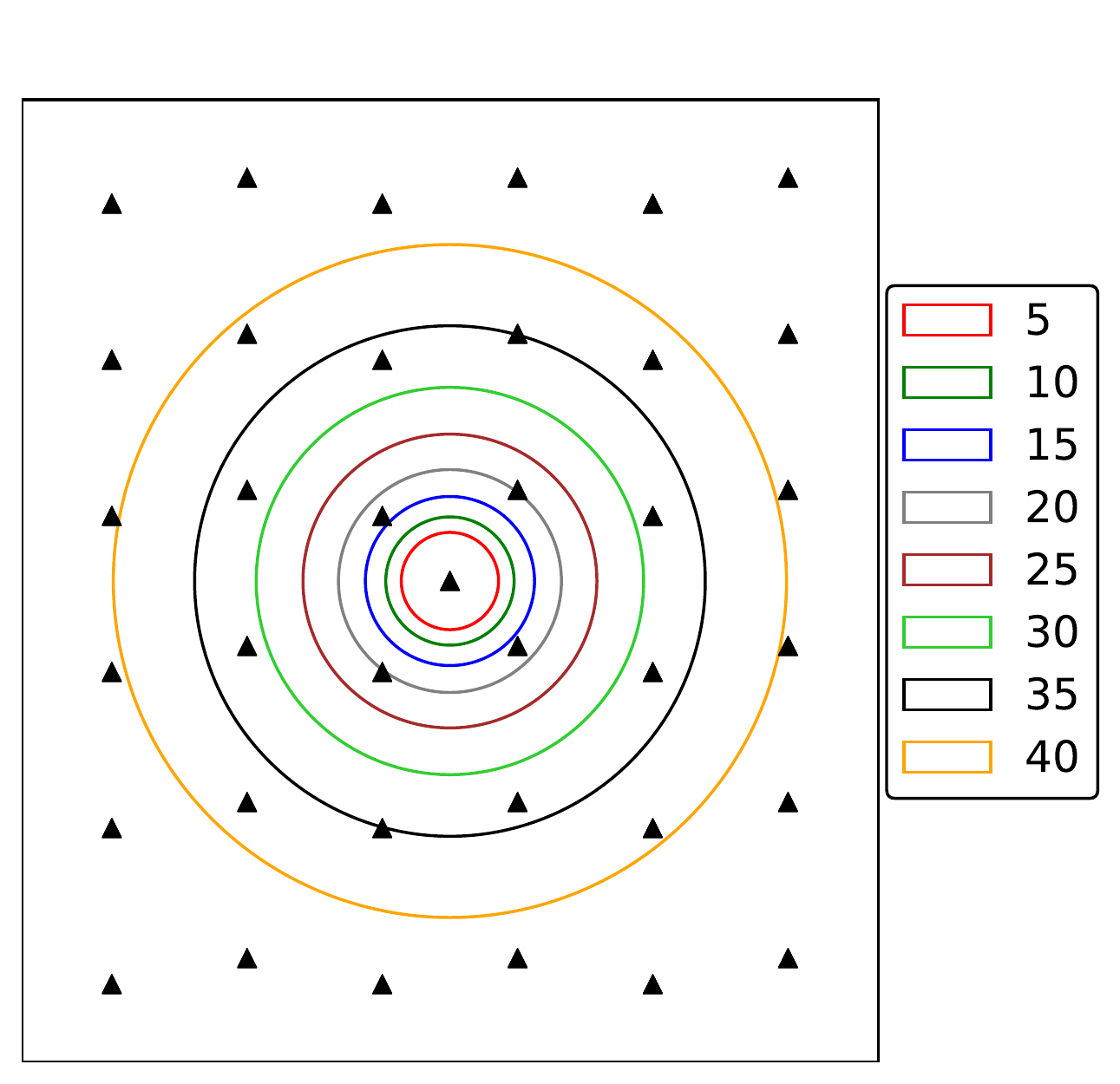}
    \caption{Reconfiguration range for variable TX Power}
    \label{slabicki:fig:range_circles}
\end{figure}

The intuitive approach to select the range of cells considered for the reconfiguration is to start with the change of the nearest neighbors and continue by increasing the range of changes. We evaluate such method: an algorithm that starts by the designation of circles with a center in the location of a newly added BS or a center of a changed state that initiates the need for reconfiguration. The set of BSes considered for a configuration change at a given step is defined by a set of a BSes located within particular circle. The radius of the smallest circle can be calculated using the free space radio signal propagation model, as a distance at which the received power of reference signal is equal to the noise level for a default TX power. The following circles are of growing size, each one with a radius multiplied by a parameter greater than 1 or can be calculated as the size of the first circle, but for a larger TX Power. A sample set of circles is presented on the Figure \ref{slabicki:fig:range_circles}. 

The base stations within the given circle define the search space for the optimization algorithm for the network reconfiguration. The configuration of all the remaining base stations is fixed and does not change at a given step. We start by running the Algorithm 1 with the change range limited to the inner circle and next we increase the range up to a point at which there is no difference between the result of the calculation for a current step and for the previous one. 

\begin{figure}
	\centering    
    \includegraphics[width=0.95\linewidth]{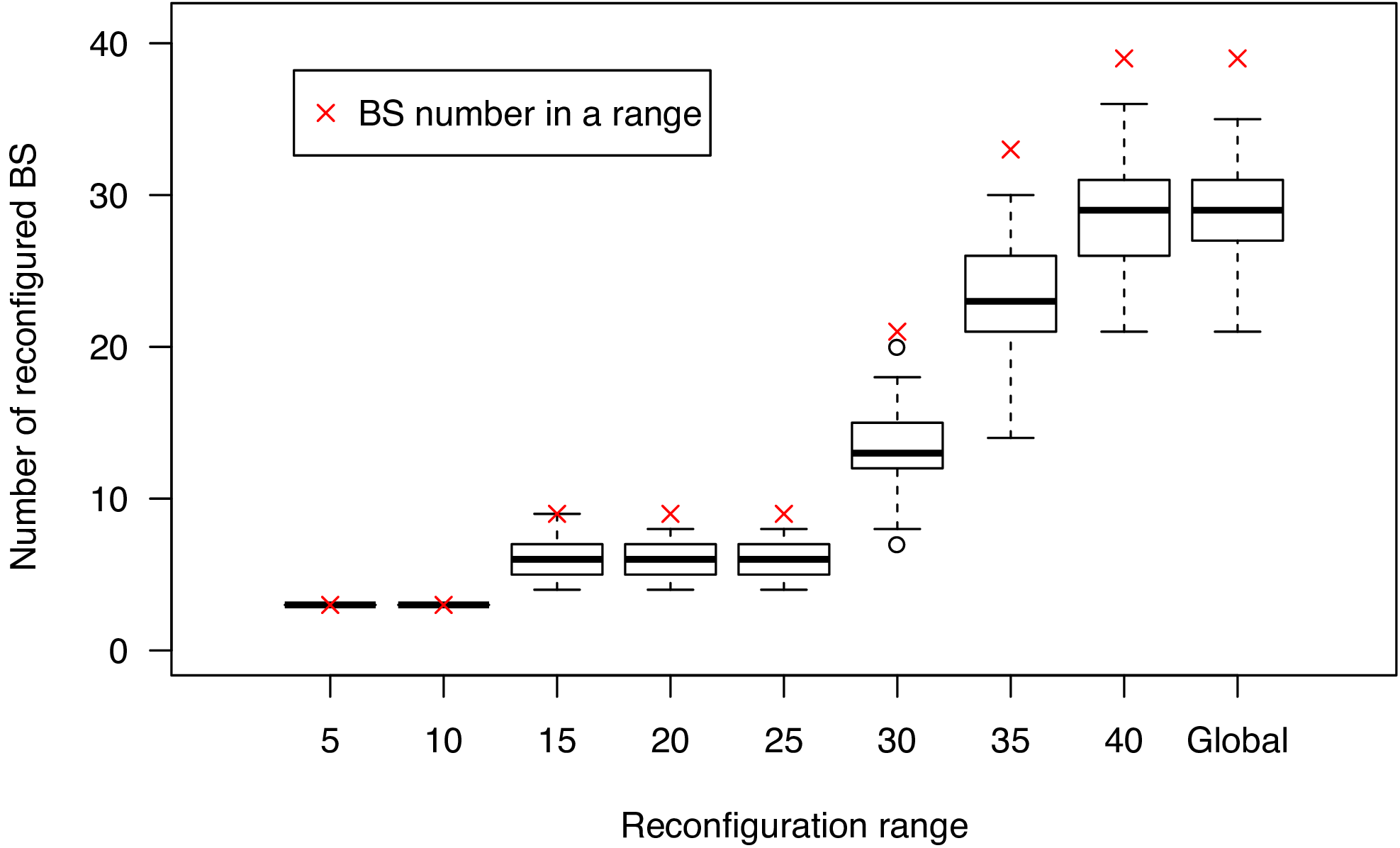}
    \caption{Number of reconfigured BSes (network of 36 cells)}
    \label{slabicki:fig:big_network_reconf_number_36}
\end{figure}

\begin{figure}
	\centering    
    \includegraphics[width=0.95\linewidth]{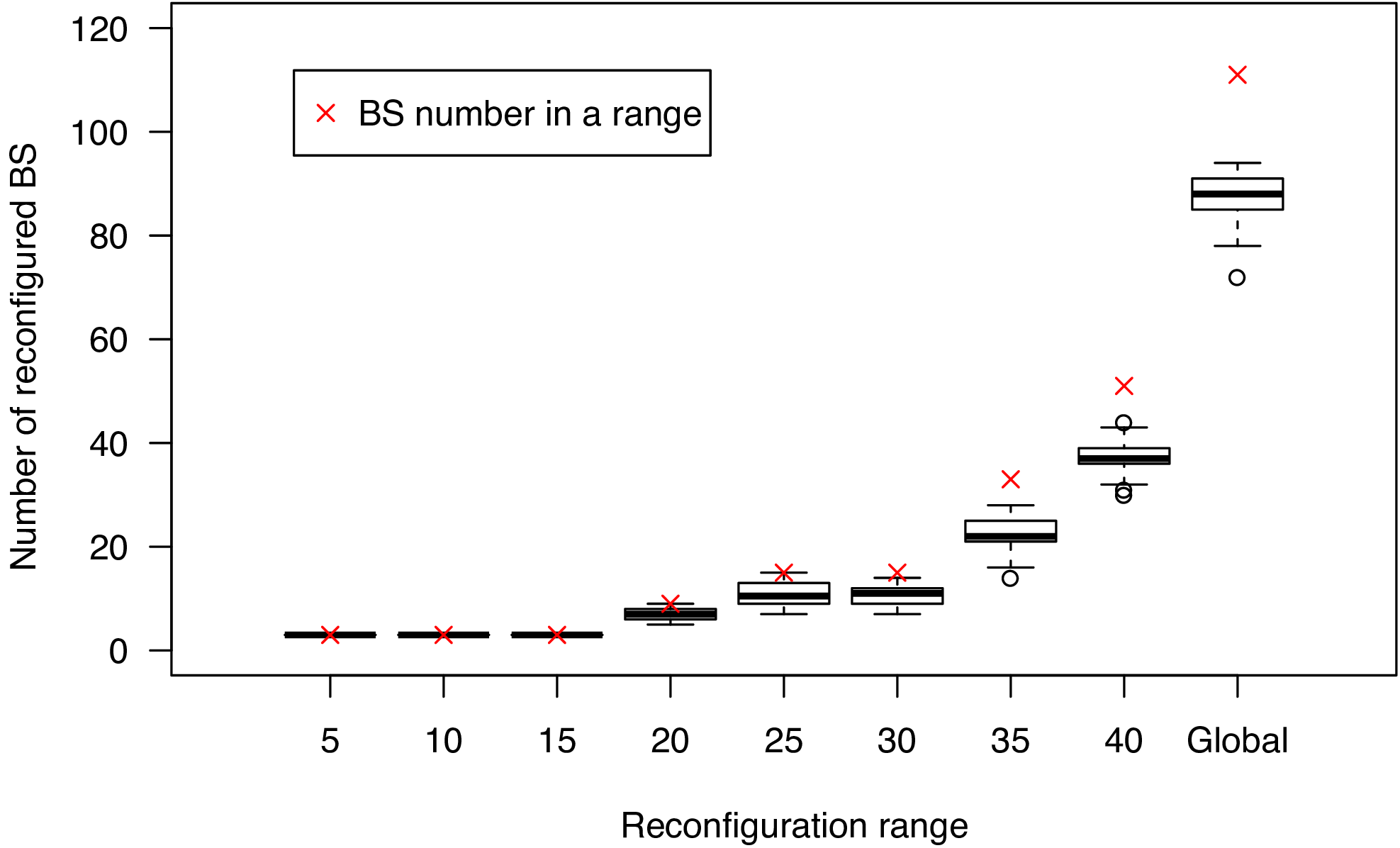}
    \caption{Number of reconfigured BSes (network of 108 cells)}
    \label{slabicki:fig:big_network_reconf_number_108}
\end{figure}

The plots \ref{slabicki:fig:big_network_reconf_number_36} and~\ref{slabicki:fig:big_network_reconf_number_108} present the box-plots of the number of reconfigured base stations for each of the sizes of the sets used for the reconfiguration. The algorithm has been executed 50 times for each set. It can be seen that in most of the cases the optimization changes the configuration only of a part of the base stations within a given circle. Additionally, the search space grows proportionally to the growing radius of the circle. 

\begin{figure}
	\centering    
    \includegraphics[width=0.95\linewidth]{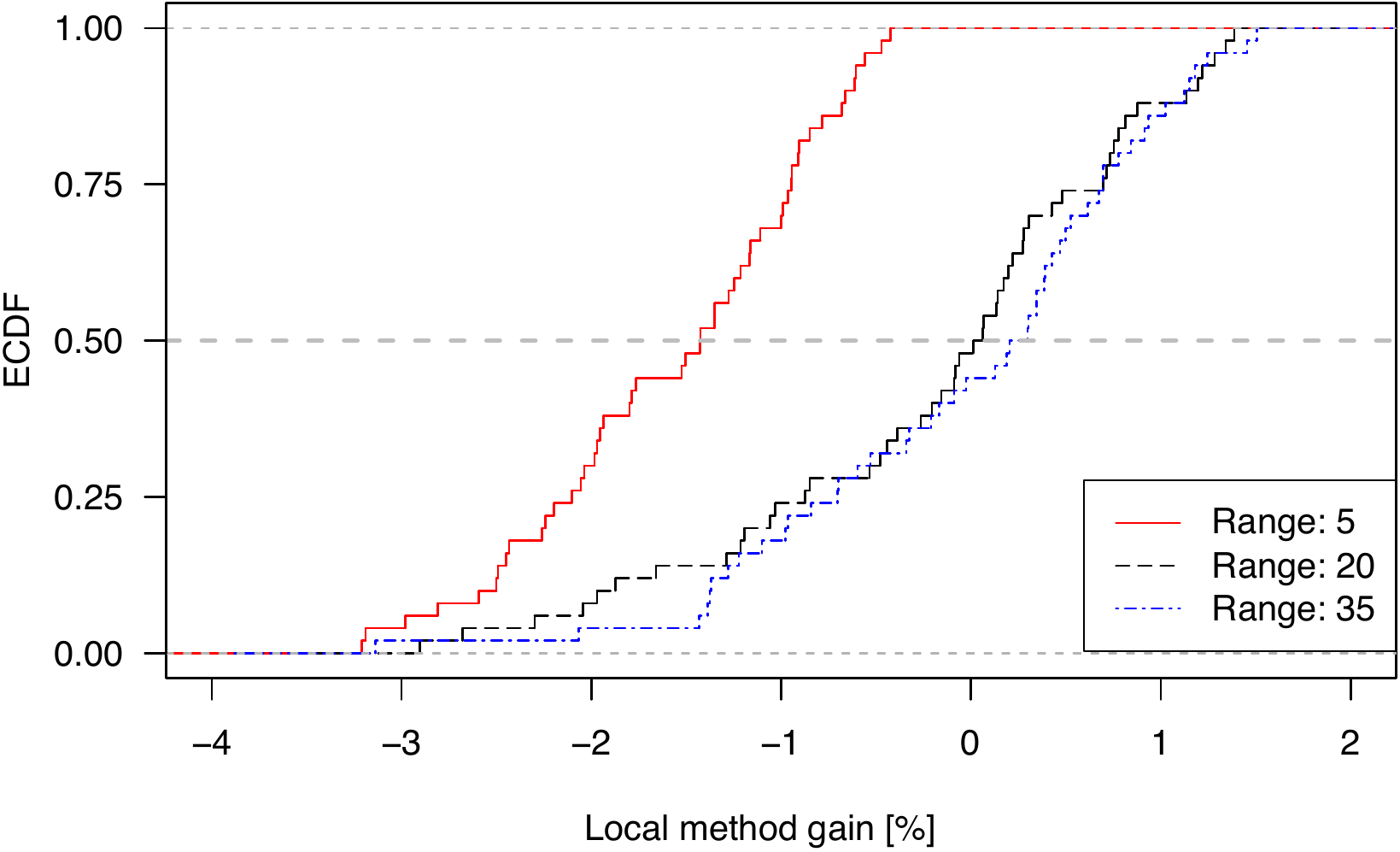}
    \caption{ECDF (network of 36 cells)}
    \label{slabicki:fig:big_network_zysk_36}
\end{figure}

\begin{figure}
	\centering    
    \includegraphics[width=0.95\linewidth]{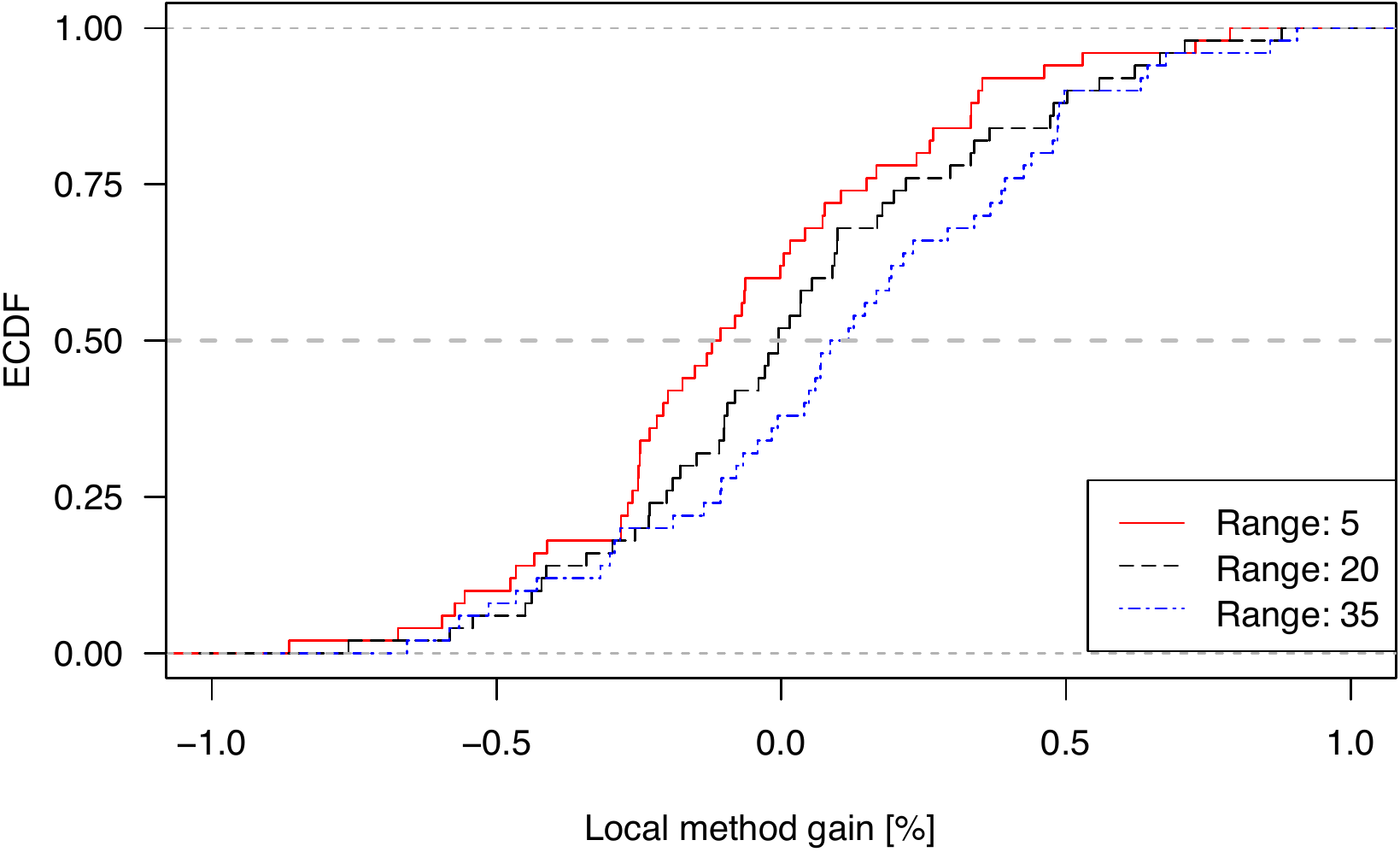}
    \caption{ECDF (network of 108 cells)}
    \label{slabicki:fig:big_network_zysk_108}
\end{figure}


Next, we compare the outcome of a various ranges of the local reconfiguration method with the results of global optimization. We have executed the following experiment: a new nomadic base station with three sectors (three cells) have been added to a network. Next we have calculated the optimized set of TX power using the global optimization and using the local reconfiguration method with three different ranges. For each of the configurations the relative difference in total network throughput. The experiment has been repeated 50 times with different random seeds. The ECDF have been plotted to present the number of experiments in which the difference between the local and global optimization. The results are presented on Figures \ref{slabicki:fig:big_network_zysk_36} and \ref{slabicki:fig:big_network_zysk_108}. For larger network (108 cells) there is practically no difference between outcome of the local reconfiguration and the global optimization of parameters. For smaller networks we can observe that the global optimization outperforms the local method by up to 2 percent, but this is the case only for a very limited range of the reconfiguration. When the range is not so limited or the network size is bigger the outcome both methods provide very similar level of the resulting total throughput of the network, but the local method reaches this goal by reconfiguring smaller number of base stations.

\section{Sampling based local TX power reconfiguration}
\label{sec:sampling}

The iterative increase of reconfiguration range method described in previous section is arbitrary in the selection of base stations which are subject of the reconfiguration. The set of base stations which are reconfigured is incrementally increased basing on the distance from the newly added cell, which is not always a adequate. In heterogeneous networks with cells of different size, some of the pico- and nano-cells could be omitted, as they will have no influence on the interferences for base stations added even in locations not so distant. Thus we have continued our work to propose a dynamic local reconfiguration method, which selects the cells which needs to be reconfigured basing on sampling. 

The goal of this algorithm is to estimate whenever the change in the network is needed basing on a limited number of sample locations, for which the received signal strength is calculated before and after the change of configuration. If the metric of quality (e.g. the average value) for the selected configuration in the points chosen is worse than the expected value, the base station is subject of reconfiguration. The algorithm omits the locations for which the influence of the configuration change is negligible or negative, e.g. because of the large distance from base station. Pseudocode is presented in Algoritm \ref{alg:sampling}.

\begin{algorithm}
\caption{Sampling based local TX power optimization}
\label{alg:sampling}
\begin{algorithmic}
\ForAll{base station}
	\State \Call{randomly select 50 pairs of coordinates (X, Y) within the base station range}{}
	\State \Call{calcualte the average SINR in selected locations}{}
	\If{average SINR after the change < threshold value} 
		Add the base station to the list
	\EndIf
\EndFor
\State \Call{Return the base stations list}{}
\end{algorithmic}
\end{algorithm}

We propose to use the threshold on average SINR change as the metric used for selection whenever the base station needs to be reconfigured. In our experiment we have used the 2dB as the threshold.


We have compared the outcome of the sampling based local TX power reconfiguration method with the global optimization method and the iterative local reconfiguration method proposed in the previous section. The analysis has been executed for three network sizes, to analyze the performance of the proposed method in networks with various density of cells and in network topology representing irregular placement of cells. The following scenarios have been implemented:

\begin{itemize}
\item 36 base stations in honeycomb topology, with simulated 324 UEs,
\item 75 base stations in honeycomb topology, with simulated 625 UEs,
\item 63 base stations in topology representing the real locations of LTE cells in Hannover city \cite{rose2013ic}, with 625 simulated UEs,

\end{itemize}

The experiments have been executed 50 times to select the configuration of each reference networks. The following three methods have been used:
\begin{itemize}
\item global optimization,
\item local reconfiguration with constant radius of the reconfiguration (described in the previous section). 
\item local with sampling.
\end{itemize}

\begin{figure}
	\centering    
    \includegraphics[width=0.95\linewidth]{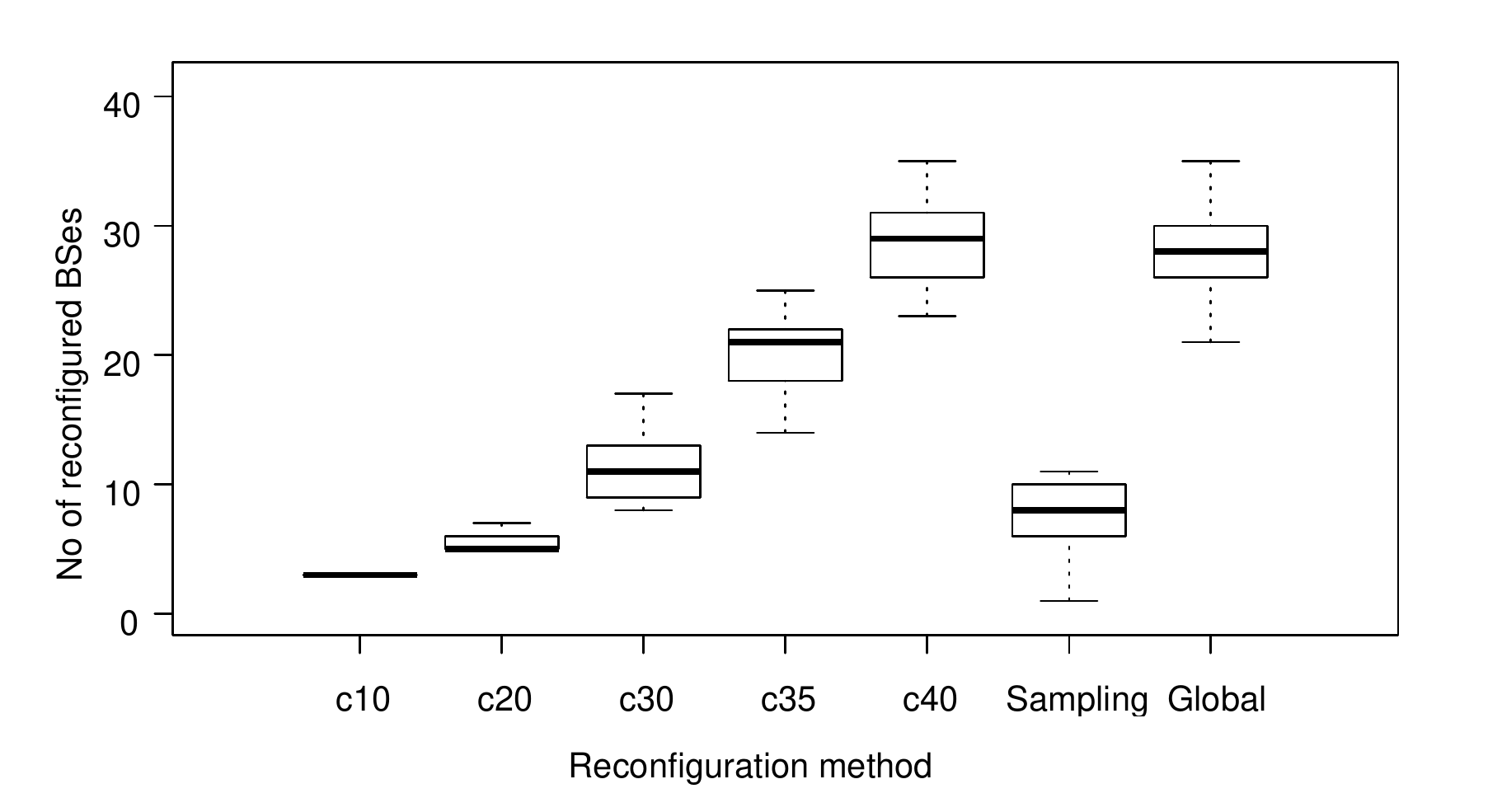}
		\caption{Boxplot showing number of reconfigured BSes for different reconfiguration methods: with range of changes limited to 10 -- 40 nearest BSes (c10 -- c40) and sampling method (36 cells, honeycomb)}
    \label{fig:sampling_bsnumber_boxplot_36}
\end{figure}

\begin{figure}
  \centering
  \includegraphics[width=0.95\linewidth]{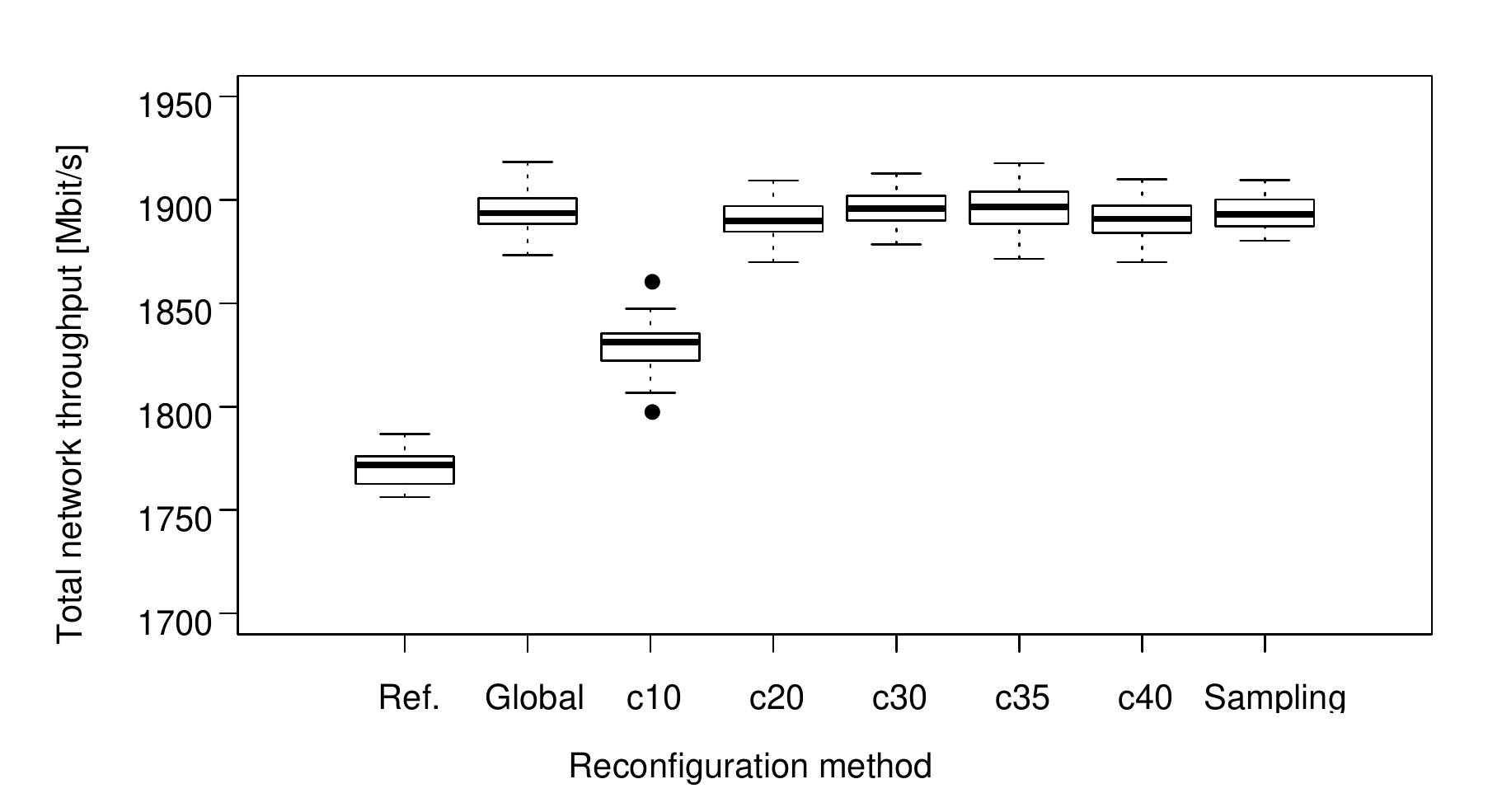}
  \caption{Boxplot showing number of reconfigured BSes for different reconfiguration methods: with range of changes limited to 10 -- 40 nearest BSes (c10 -- c40) and sampling method (36 cells, honeycomb)}
  \label{fig:sampling_thr_boxplot_36}
\end{figure}

The box plot on Figure \ref{fig:sampling_bsnumber_boxplot_36} shows the relation between the number of reconfigured base station in response to the addition of a new cell in the network consisting 36 base stations. The sampling local reconfiguration algorithm requires to make lower number of changes in configuration (the median is 10) even comparing to the iterative approach with a small radius, but provides the same outcome in terms of the total offered throughput in the network, which is presented on the Figure \ref{fig:sampling_thr_boxplot_36}.

\begin{figure}
	\centering    
    \includegraphics[width=0.95\linewidth]{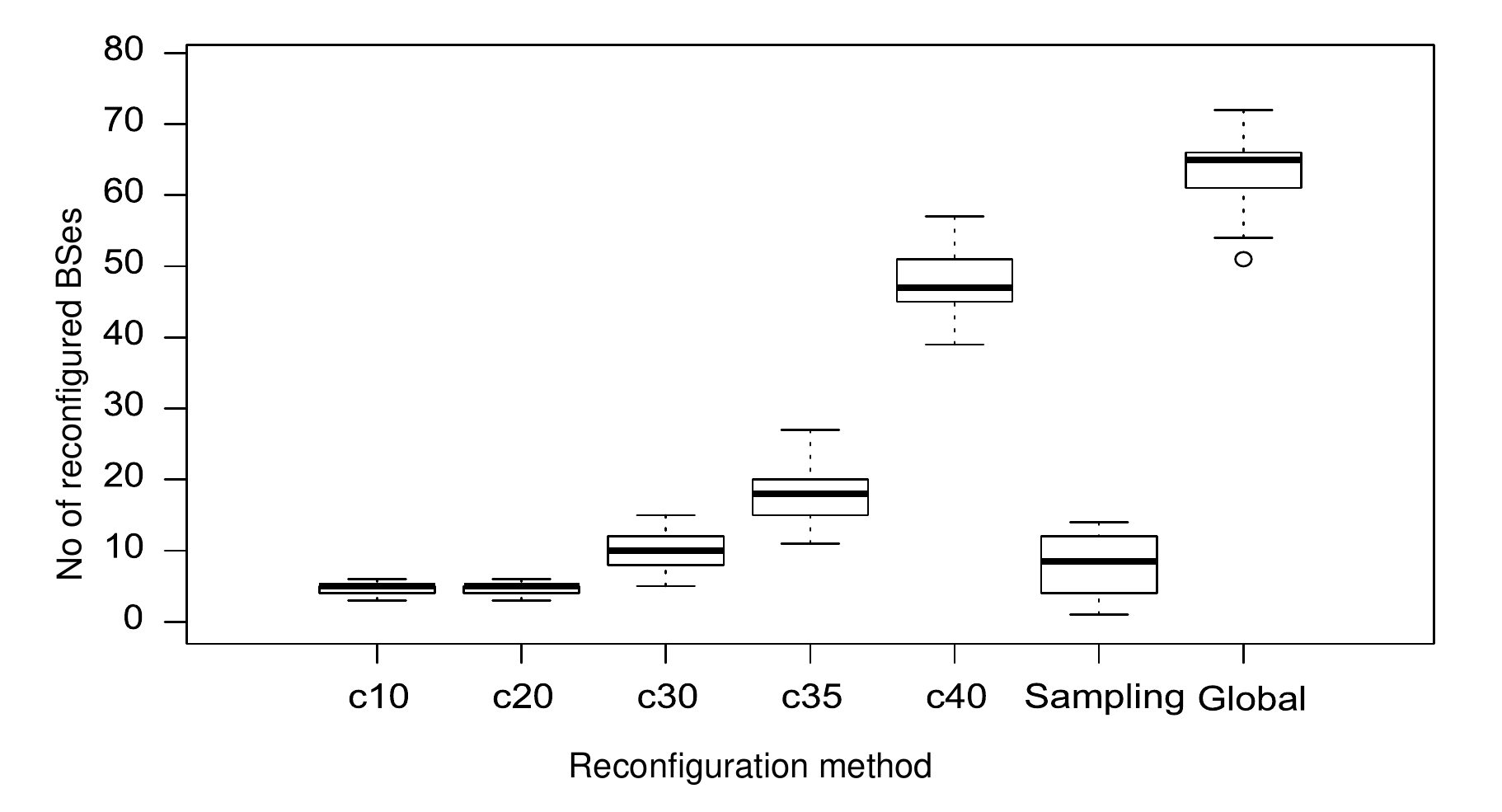}
		\caption{Boxplot showing number of reconfigured BSes for different reconfiguration methods: with range of changes limited to 10 -- 40 nearest BSes (c10 -- c40) and sampling method (75 cells, honeycomb)}
    \label{fig:sampling_bsnumber_boxplot_75}
\end{figure}
\begin{figure}
  \centering
  \includegraphics[width=0.95\linewidth]{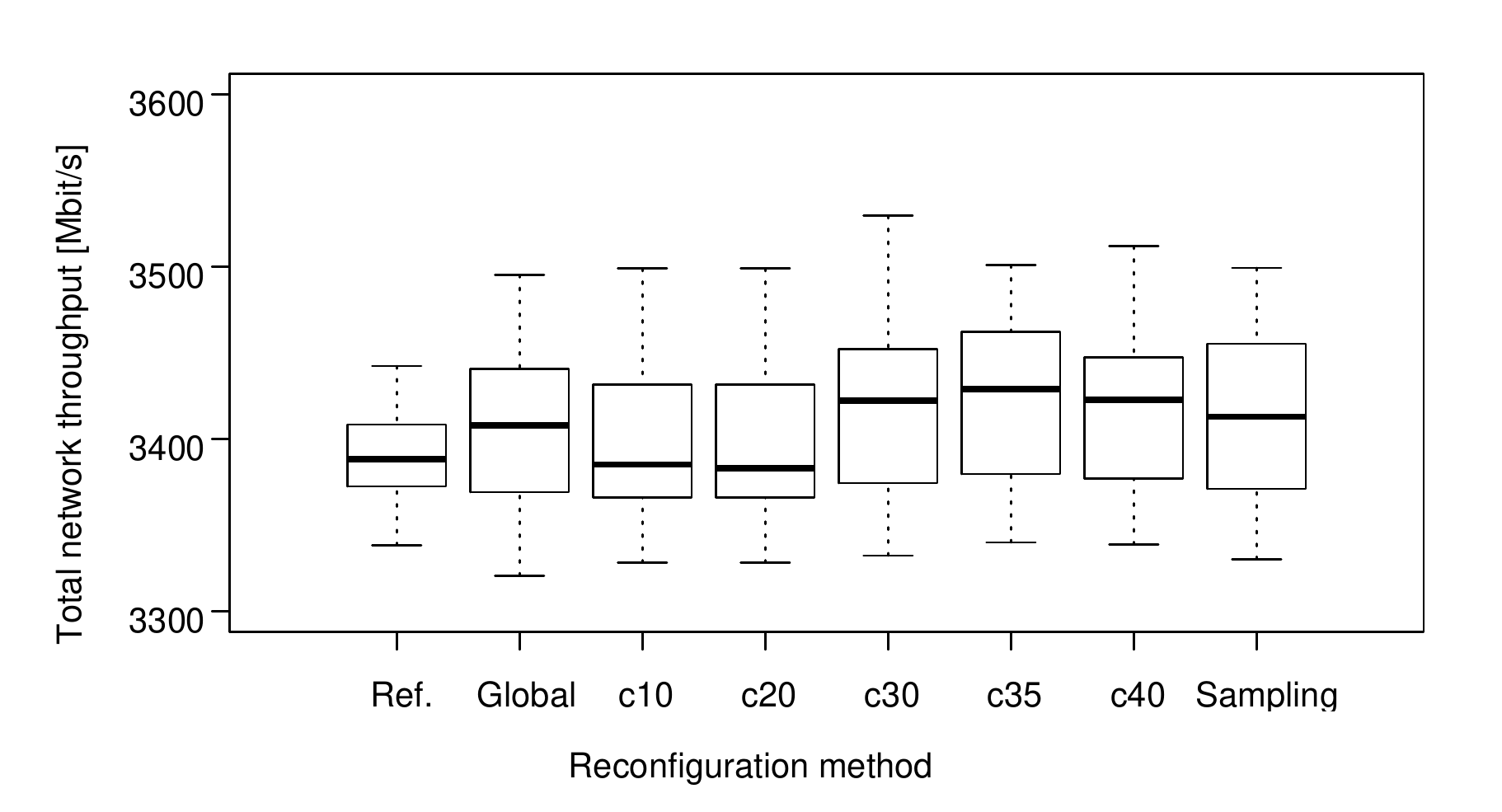}
  \caption{Boxplot of total throughput for different reconfiguration methods (75 cells, honeycomb)}
  \label{fig:sampling_thr_boxplot_75}
\end{figure}
\begin{figure}
	\centering    
    \includegraphics[width=0.95\linewidth]{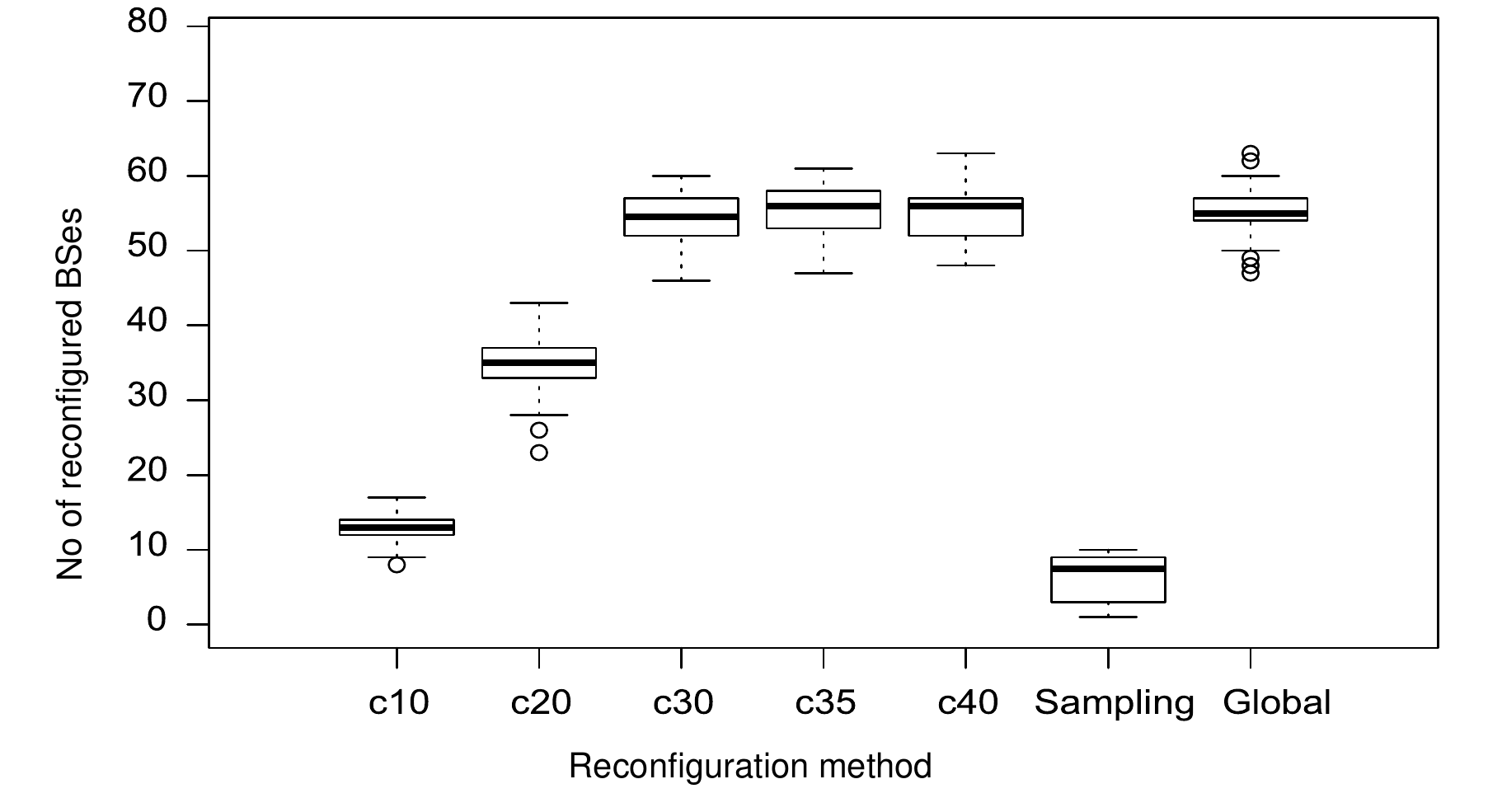}
		\caption{Boxplot showing number of reconfigured BSes for different reconfiguration methods (63 cells, Hannover topology)}
    \label{fig:sampling_bsnumber_boxplot_han}
\end{figure}
\begin{figure}
  \centering
  \includegraphics[width=\linewidth]{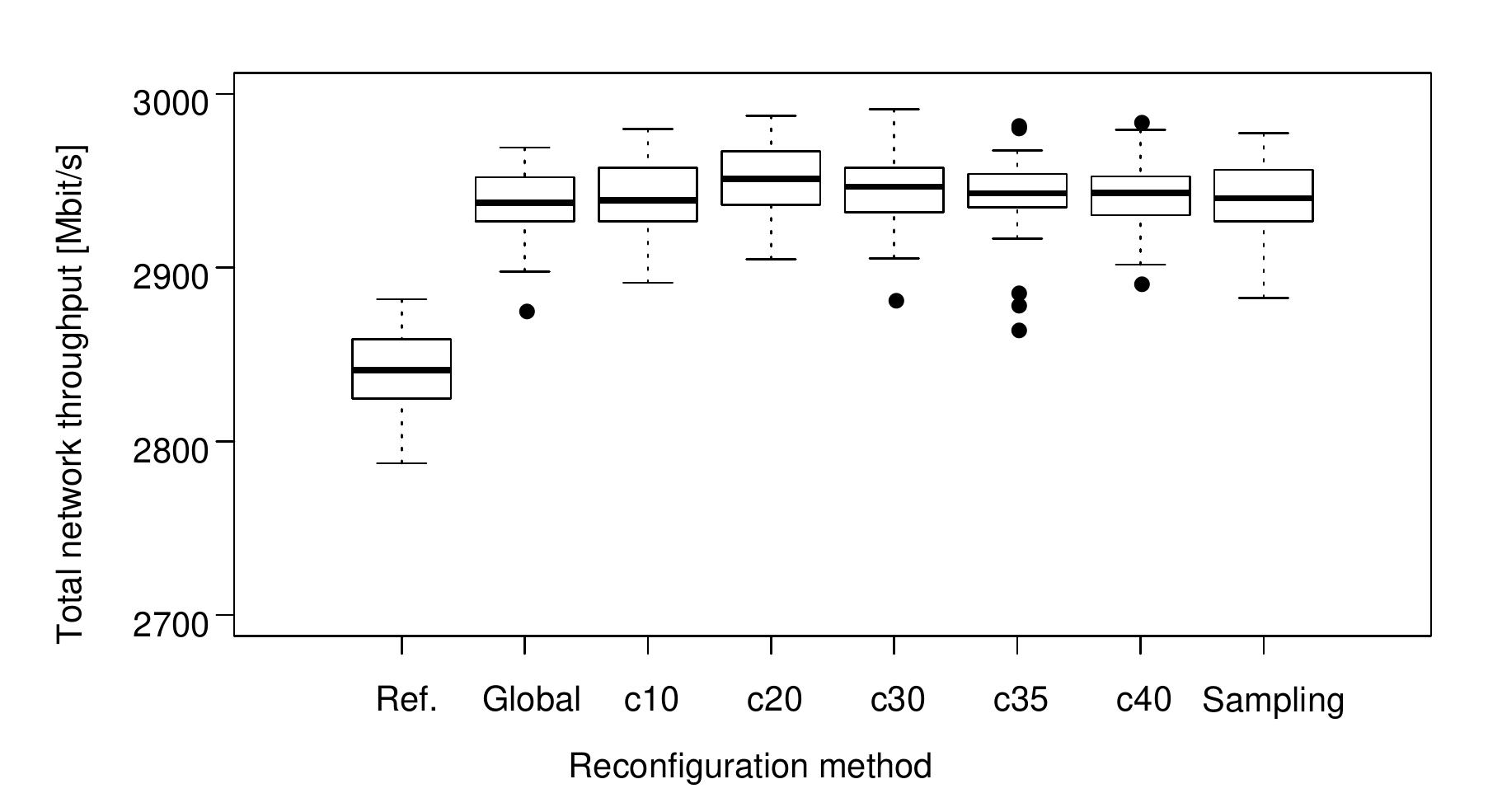}
  \caption{Boxplot of total throughput for different reconfiguration methods (63 cells, Hannover topology)}
  \label{fig:sampling_thr_boxplot_han}
\end{figure}

The analysis of the sampling algorithm in larger network (75 base stations) shows similar results, as can be observed on the Figures \ref{fig:sampling_bsnumber_boxplot_75} and \ref{fig:sampling_thr_boxplot_75}. While the number of reconfigured stations is slightly higher because of the larger size of the network, the proportion remains similar and the sampling algorithm is able to provide results as good as the global optimization, with significantly lower amount of reconfigurations needed. The results on Hannover city network topology confirm that the sampling algorithm works well in the real-life structure of the network, and the Figures \ref{fig:sampling_bsnumber_boxplot_han} and~\ref{fig:sampling_thr_boxplot_han} show that in network with variable sizes of the cells it is able to provide throughput as good as global optimization, requiring to reconfigure less than 10 base stations. 


\section{Conclusions}
\label{sec:conclusions}

The local approach to reconfiguration of the network, in response to the addition of a new cell to existing network, allows to limit the number of configuration changes needed to adapt the transmission power to the new network topology. This work shows how to select the range for TX Power reconfiguration when a new cell is added to the network. We propose sampling based algorithm that allow to select the set of cells which needs to be reconfigured. The analysis of the proposed method in regular topology and in real life scenario of base stations located in a city environment shows that it provides similar level of throughput offered to the clients comparing to the reconfiguration of the whole network based on global optimization, through a reconfiguration of only a small set of base stations. The sampling based algorithm is able to reach the near optimal configuration with smaller number of reconfigurations: by limiting the range of changes to on average 10 neighboring cells, the achieved efficiency of the network is the same as with global optimization of TX power. This allows 5G and LTE operators to react faster and with lower cost to the changes in network and events such as addition of a new nomadic base station or deployment of a small cell.

\bibliography{ms_phd_bib}{}
\bibliographystyle{plain}

\end{document}